# Deep Learning Bandgaps of Topologically Doped Graphene

Yuan Dong[1ξ], Chuhan Wu[2ξ], Chi Zhang[1], Yingda Liu[2], Jianlin Cheng[2*], Jian Lin[1*]

[1]Department of Mechanical & Aerospace Engineering, University of Missouri, Columbia, Missouri 65211, USA
[2]Department of Electrical Engineering & Computer Science, University of Missouri, Columbia, Missouri 65211, USA

[*]E-mail: LinJian@missouri.edu (J. L.) or chengji@missouri.edu (J. C.)
[ξ]Authors contributed equally to this work.

**Abstract**

Manipulation of material properties via precise doping affords enormous tunable phenomena to explore. Recent advance shows that in the atomic and nano scales topological states of dopants play crucial roles in determining their properties. However, such determination is largely unknown due to the incredible size of topological states. Here, we present a case study of developing deep learning algorithms to predict bandgaps of boron-nitrogen pair doped graphene with arbitrary dopant topologies. A material descriptor system that enables to correlate structures with the bandgaps was developed for convolutional neuron networks (CNNs). Bandgaps calculated by the *ab initio* calculations and the corresponding structures were fed as input datasets to train VGG16 convolutional network, residual convolutional network, and concatenate convolutional network. Then these trained CNNs were used to predict bandgaps of doped graphene with various dopant topologies. All of them afford great prediction accuracy, showing square of the coefficient of correlation ($R^2$) of > 90% and root-mean-square errors of ~ 0.1 eV for the predicted bandgaps. They are much better than those predicted by a shallow machine learning method - support vector machine. The transfer learning was further performed by leveraging data generated from smaller systems to improve the prediction for large systems. Success of this work provides a cornerstone for future investigation of topologically doped graphene and other 2D materials. Moreover, given ubiquitous existence of topologies in materials, this work will stimulate widespread interests in applying deep learning algorithms to topological design of materials crossing atomic, nano-, meso-, and macro- scales.

**Keywords:** 2D materials, bandgaps, convolutional neural networks, deep learning, DFT

# 1. Introduction

Introduction of defects at atomic or nanoscales has been a widely employed strategy in bulk materials, such as metals, ceramics, and semiconductors, for manipulation of their mechanical and physical properties. Especially when these defects are precisely controlled in topological space, they significantly improve the mechanical properties,(1-3) manipulate the magnetic properties,(4-6) and alter the electronic properties.(7-10) For instance, nanotwined grain boundaries are crucial to realize ultrahigh strength, superior fatigue resistance in metals,(1) and ultrahigh hardness and enhanced toughness in ceramics.(2, 3) Localization of nitrogen vacancy center at deterministic positions with nanoscale precision within diamond improves sensitivity and resolution of single electron spin imaging,(4, 5) and enhances charge-state switching rates,(6) paving new ways to next-generation quantum devices. Precisely controlling dopant atoms in term of concentration and spatial arrangement within semiconductors is so critical to performance of fabricated electronic devices, especially as the dimension of the device keeps shrinking.(7) Manipulation and detection of individual surface vacancies in a chlorine terminated Cu (100) surface realizes atomic scale memory devices.(10)

Based on the aforementioned examples in the bulk materials, it is naturally anticipated that the effect of the defects in two-dimensional (2D) materials, in term of the defect density and topologies, would be even more profound as all the atoms are confined within a basal plane with atomic thickness. Distinguished from allowance of different pathways of defect configurations in the three-dimensional (3D) bulk materials, this dimensionality restriction in the 2D materials largely reduces the accessibility and variability of the defects. This unique topology would allow topological design of the defects in 2D, which starts to emerge as a new and promising research field. As the first well-known 2D material, graphene has been shown to exhibit topological grain boundaries-dependent mechanical,(11, 12) thermal,(13) and electrical properties.(14, 15) Doping graphene with heteroatoms, such as hydrogen, nitrogen, and boron, can further tailor the magnetic or electrical properties.(16) Theoretical calculations suggest that these properties not only depend on types of dopants,(16-18) and doping concentrations,(19) but also are greatly determined by the dopants topologies within the graphene.(20-22) Although the theoretical investigation already enables to research a much larger set of cases than the experiment does. The number of possible topological configurations for the dopants in graphene far exceeds the amount that can be practically computed due to extremely high computational cost. For instance, doping boron-nitrogen (B-N) pairs into a graphene layer with a just $6 \times 6$ supercell system results in billions of possible topologically doped graphene structures. Thus, it is entirely impractical to study all the possibilities to get the optimized properties even for such a small system. Another limitation of current mainstream of material design is that it heavily relies on intuition and knowledge of human who design, implement, and characterize materials through the trial and error processes.

Recent progress in data-driven machine learning (ML) starts to stimulate great interests in material fields. For instance, a series of material properties of stoichiometric inorganic crystalline materials were predicted by the ML.(23) In addition to their potentials in predicting properties of the materials, they start to show great power in assisting materials design and synthesis.(24-27) It is anticipated that ML would assist to push the material revolution to a paradigm of full autonomy in the next five to ten years,(28, 29) especially as emerging of deep learning (DL) algorithms.(30, 31) For instance, a pioneer work of employing only a few layered convolutional neural networks (CNNs) enables to reproduce the phase transition of matters.(32) Nevertheless, the application of the DL in the material fields is still in its infancy.(33) One of main barriers is that a compatible and sophisticated descriptive system that enables to correlate the predicted properties to structures is required for materials because DL algorithms are originally developed for imaging recognition.

Motivated by this challenge, we conceive to employ CNNs, including VGG16 convolutional network (VCN), residual convolutional network (RCN), and our newly-constructed concatenate convolutional network (CCN), for predicting electronic properties of graphene doped with randomly configured boron-nitrogen (B-N) pairs. As a benchmark comparison, the support vector machine (SVM),(34) which used to be the mainstream ML algorithm before the DL era, was also adopted (see details in Supplementary Note 1). We discovered that after trained with structural information and the bandgaps calculated from *ab initio* density function theory (DFT), these CNNs enabled to precisely predict the bandgaps of B-N pair doped graphene with any given dopant topologies. One main reason for the high prediction accuracy arises from the developed material descriptor which enables to extract the features of structures. In the descriptive system, we set "1" for a B-N pair while "0" for a carbon-carbon pair. It is a natural representation like a 2D image, which is very suitable for the DL algorithms. Moreover, this material descriptor well matches the bandgap distribution of graphene and BN mixtures. Because the bandgap of intrinsic BN is 4.588 eV while the intrinsic graphene has a bandgap of 0 eV. Mixing of them would show a certain relation, but may not be simply linear. The convolutions in the CNNs can extract the features not only from the elements of the input data but also from their neighbors. Such a descriptive system enables to qualitatively and quantitatively capture the features of topological doping states, where each atom in the structure affects its neighbor atoms so that these localized atomic clusters collectively determine bandgaps of the whole structure. Combined with well-tuned hyperparameters and well-designed layers, the CNNs result in high prediction accuracy. Considering that atom-scale precise structures of doped graphene by bottom-up chemical synthesis have been experimental realized,(35-37) this work provides a cornerstone for future investigation of topological doping in graphene and other 2D materials as well as their associated properties. We believe that this work will bring up broader interests in applying the designed descriptive system and the CNN models for many materials related problems, which are not accessible to other machine learning algorithms.

## 2. Results

**2.1 Dataset for bandgaps and structures of topologically doped graphene**

Graphene and h-BN have similar honeycomb structures with very close bond lengths (1.42 Å and 1.44 Å for graphene and h-BN, respectively),(38) which is beneficial to structural stability if B-N pairs are doped in graphene. Moreover, graphene is a semimetal with a zero bandgap while h-BN is a wide-bandgap semiconductor. Thus, it can be naturally assumed that the graphene doped with the B-N pairs could have an intermediate bandgap. Finally, the B-N pairs can exactly make the charge neutral in the doped graphene. To test these hypotheses, we first implemented high throughput DFT calculations on topologically doped graphene to generate datasets—correlation of chemical structures to bandgaps—for the ML. To achieve high-throughput calculations, we only apply non-hybrid function for the DFT calculations. Even though the calculated bandgaps are not as accurate as the previously reported values calculated by the hybrid functions, data consistence can be secured. Initial calculations show that the bandgaps of the pristine graphene with $4 \times 4$, $5 \times 5$, and $6 \times 6$ supercell systems are exactly 0 eV (Fig. 1A), and the bandgap of the pristine h-BN is 4.588 eV (Fig. 1B). More DFT calculations on examples of $4 \times 4$ systems that have the 50 at.% B-N dopant concentrations but with different topological states show that they exhibit bandgaps ranging from 0.9525 to 1.5705 eV (Fig. 1C). As the dopant concentration increases to 62.5 at.%, the corresponding bandgaps also increase (Fig. 1D). These results validate that bandgaps of doped graphene depend on the dopant topologies. In addition, we also compared the DFT results with the experimental values. Ci et al. reported the hybrid BNC sheets with bandgaps of 1.62 eV and 1.51 eV for the samples with 35 at.% B-N and 16 at.% B-N dopant concentrations.(39) In our calculation, the 5 ×5 systems with 36 at.% B-N dopants have the bandgaps ranging from 0.75 eV to 1.31 eV, and the systems with 16 at.% B-N dopants have the bandgaps ranging from 0.45 eV to 0.61 eV. Although they are smaller than the experimental values, they are still in an acceptable discrepancy range. Because in experiment, the BN dopants tends to form large clusters, which would increase the bandgaps of these hybrid atomic sheets.(39, 40) Finally, more bandgaps of doped graphene with arbitrary dopant concentration and random dopant topologies were calculated by the DFT. Some of them served as the training datasets for training the CNNs. Others served as test datasets to validate the accuracy of the prediction performed by the CNNs.

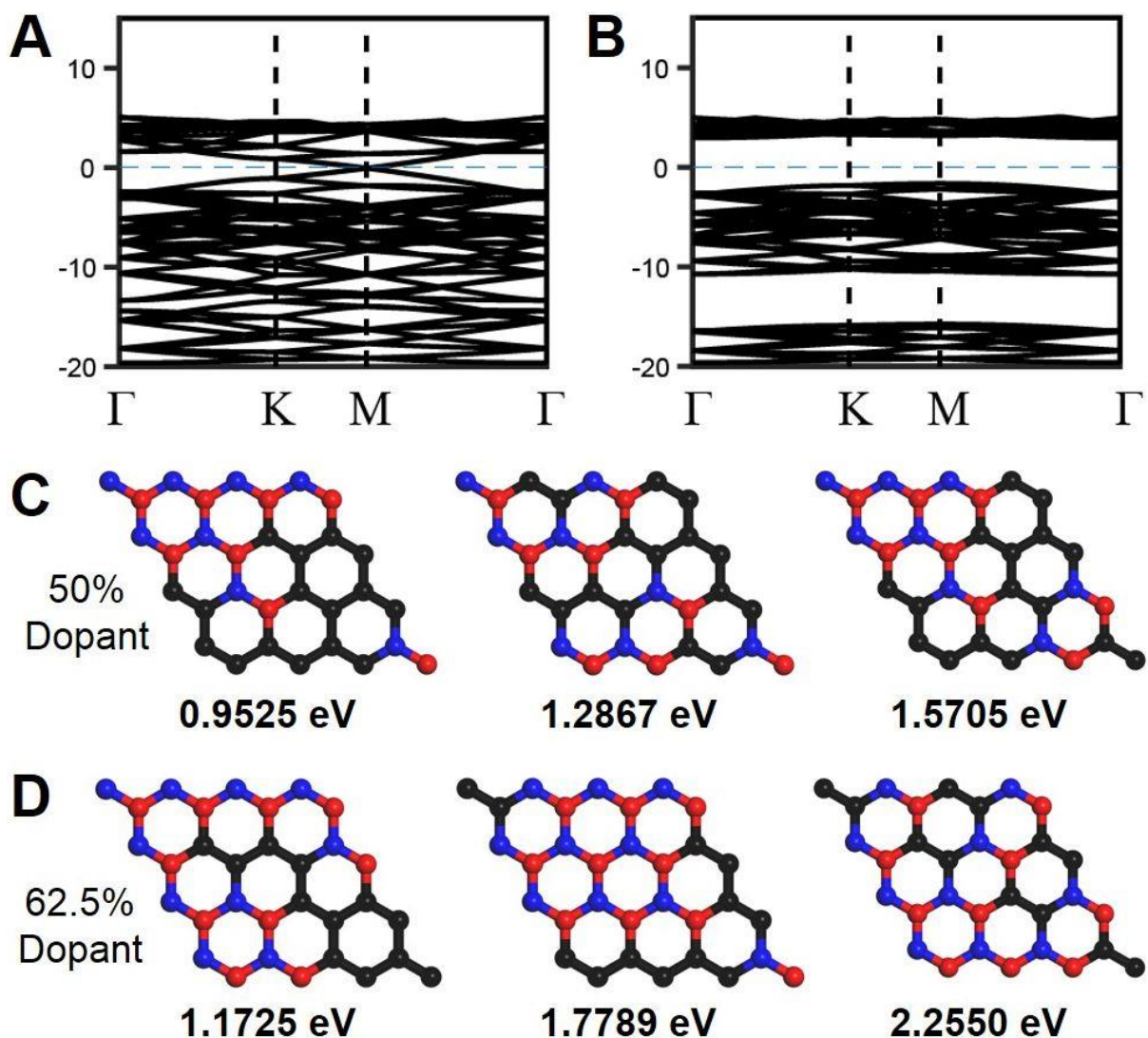

**Figure 1.** (**A**, **B**) Band structures of pristine graphene (**A**) and pristine h-BN (**B**). (**C**, **D**) Representative atomic configurations of 4 ×4 graphene supercell systems that have the same B-N dopant concentrations (50% and 62.5%) but with different topological states and their corresponding bandgaps. C, N, B atoms are colored with black, blue, and red.

After the bandgap matrix is generated, structural information, such as chemical compositions and structures, needs to be well described in a form of numerical matrixes which will serve input data to train the CNNs. Defining descriptors of the materials for the ML is one of main challenges because the descriptors are more important to influence model accuracy than the ML algorithms do.(23, 41) Herein, we chose a simple and illustrative descriptor which is very suitable for the DL framework. We define the doped graphene structure by a 2D matrix described with only "0" and "1". "0" corresponds to a C-C pair while "1" corresponds to a B-N pair (Fig. 2A). The size of matrix is $4 \times 4$, $5 \times 5$, and $6 \times 6$, the same to size of supercells. This natural representation like a 2D image significantly simplifies the learning process, securing the model accuracy. Note that we exclude the cases of switching configurations of B-N pairs, which will add more complexity to the investigated systems. The size of a matrix is determined by the size of a supercell in an investigated system. For instance, a $4 \times 4$ system was represented by a $4 \times 4$ matrix. In this work, $4 \times 4$, $5 \times 5$, and $6 \times 6$ systems were studied. To enlarge the training dataset, the equivalent structures were obtained by translating the particular structures along their lattice axis or inversion around their symmetry axis (Fig. S1). These structures are equivalent, thus having the same bandgaps. By this way, ~ 14,000, ~49,000, and ~7,200 data examples for the $4 \times 4$, $5 \times 5$, and $6 \times 6$ supercell systems, respectively, were generated for purposes of training and validating CNNs. These datasets cover 21.36%, 0.15%, and $1 \times 10^{-7}$ of all possible dopant topologies for the $4 \times 4$, $5 \times 5$, and $6 \times 6$ supercell systems, respectively.

**2.2 Construction of convolutional neural networks**

The general procedure of setting up the CNNs is illustrated as follows. The structural information of the doped graphene sheets is represented by input matrices. The input matrices are transformed into multiple feature maps by filters in the first convolutional layer. The output of the convolutional layer is further transformed into high-level feature maps by the next convolutional layer. The maximum value of each feature map of the last convolutional layer is pooled together by the max pooling, which is used as the input of the next three fully connected (FC) layers. The single node in the output layer takes the output of the last fully connected layer as input to predict the bandgaps. The CNNs used in this work were constructed into three different structures. Firstly, we constructed a network which is similar to the traditional VCN designed for image processing,(42) as shown in Fig. 2B. This network has 12 convolutional (Conv) layers, one global-flatten layer, three full-connected (FC) layers, and an output layer. The neural layers in VCN are explained in Supplementary Note 2. The detailed hyperparameters are given in Table S1.

Although the VCN is capable of learning from the data, our prediction results show that its performance is limited by the depth of neurons. The accuracy gets saturated and degraded rapidly. To tackle this problem, we further modified VCN and developed the other two novel networks, RCN and CCN. The construction of RCN is explained in Supplementary Note 3. Its structure and hyperparameters are shown in Fig. S2 and Table S2-S4. The structure of RCN that we used is similar to that of ResNet50 network,(43) but with the Max-Pooling layer replaced with a Global Max pooling layer. A characteristic of this network is that the dimension of the data is kept the same until the last FC layer. The CCN used in this study is our newly-constructed network which combines advantages of both GoogleNet(44) and DenseNet.(45) The structure and hyperparameters of this network is explained in Supplementary Note 4 and illustrated in Fig. S3A and Table S5-S6. Unlike RCN which "adds" feature maps in element-wise, CCN concatenates the layer from input and output data by simultaneously passing them through the activation layer. The concatenation of the network prevents it from degradation which is usually existing in the VCN. They can gather evidences and information to improve its analyzing and cognitive capability. Similar to RCN, we also introduce the concatenation blocks into CCN as the building blocks (Fig. S3B and Table S6).

After the networks are built, they were trained by the generated datasets which correlate the structural information of the doped graphene with their corresponding bandgaps. These original datasets were split into the training and test datasets, respectively. For example, when investigating 4 ×4 supercell systems, we randomly chose 13,000 and 1,000 different data points from all the data examples to be training datasets and test datasets, respectively, to train and validate the deep learning methods. After the CNNs were trained, 300 new datasets that illustrate 300 types of the doped graphene structures for each type of investigated supercell systems (4 ×4, 5 ×5, and 6 ×6) were fed into the trained CNNs to predict the bandgaps. Note that these used structures do not exist in ones used for training and validation. The predicted bandgaps were compared to the values calculated by the DFT on the same structures for evaluating the prediction accuracy of the DL algorithms.

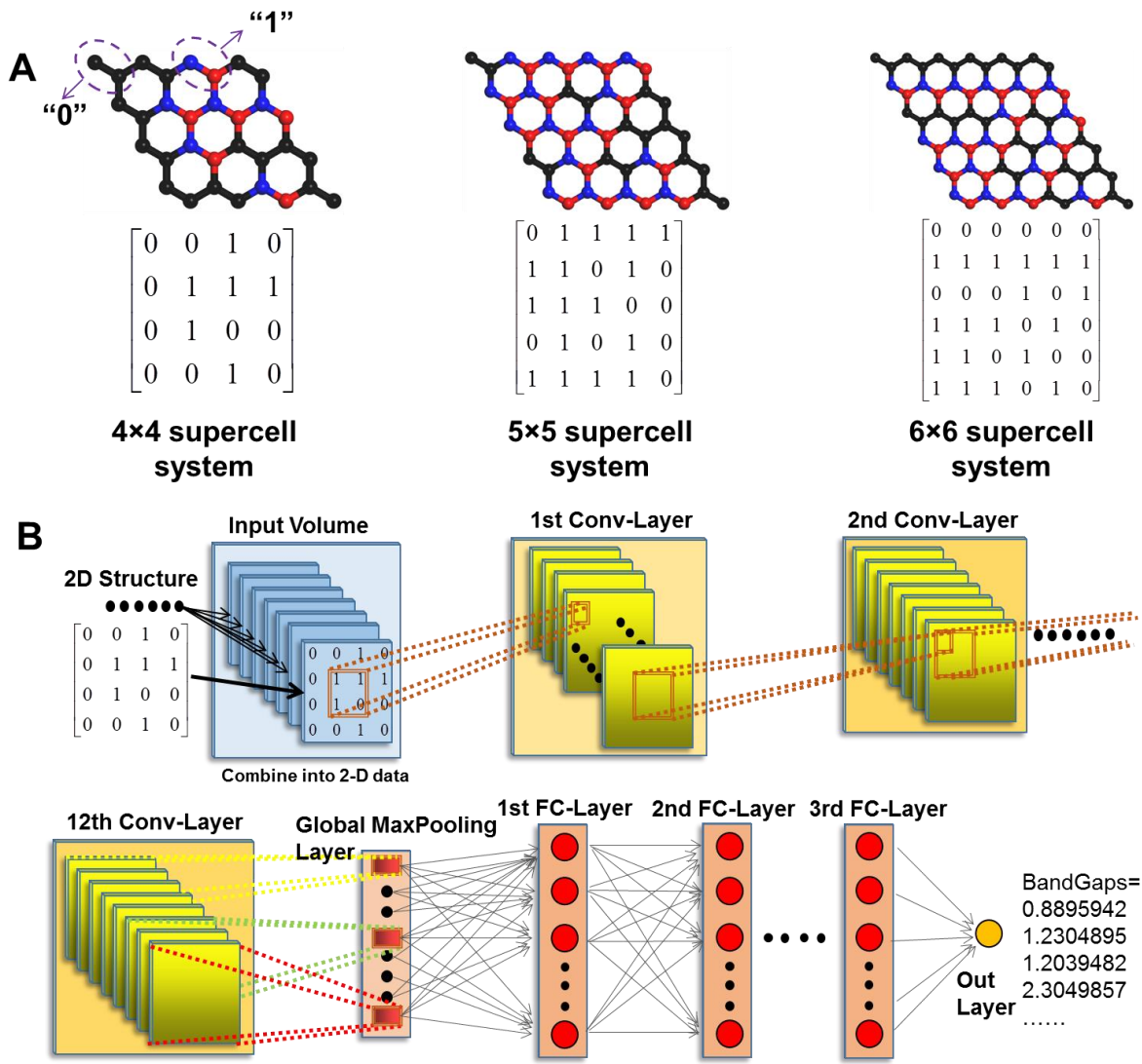

**Figure 2.** (**A**) Descriptors for 2D doped graphene supercell systems (4 × 4, 5 × 5, and 6 × 6 systems). (**B**) A convolutional neural network, VCN, for the prediction of bandgaps of 2D doped graphene systems.

## 2.3 Prediction of bandgaps by CNNs

As shown in Fig. 1, we can conclude that the bandgaps of the doped graphene are influenced by both the dopants and their topologies. In other words, each atom in the structure affects its neighbor atoms so that these localized atomic clusters collectively determine bandgaps of the whole structure. As convolutions in the CNNs can extract the features not only from the elements of the input data but also from their neighbors, it qualitatively and quantitatively captures the features of topologically doped graphene, which will be proved as follows. The predicted bandgaps of the 4 ×4 and 5 ×5 supercell systems by different DL algorithms are compared with the results by the DFT calculations (Fig. 3). Note that these data is obtained from the method of "learning from scratch" which suggests that the training and prediction are performed under the same graphene-h-BN hybrid systems.(46) For instance, the networks which are trained using the data from the 4 ×4 supercell systems are used to predict the systems with the same size but different topological configurations. The prediction accuracy is characterized by the relative error of the predicted bandgaps ($E_{ML}$) to the DFT calculated bandgaps ($E_{DFT}$), which is calculated as $|E_{ML}-E_{DFT}|/E_{DFT}$. As shown in Fig. 3A, all of three CNNs can predict the bandgaps of 4 ×4 supercell systems within 10% relative error for > 90% cases. All of the predicted bandgaps for all cases have accuracy of > 80%. In contrast, the prediction results from the SVM are deviated much more from the DFT benchmarks, showing > 20% error for > 50% cases. Fig. 3B shows that the three CNNs exhibit strong direct linear correlation of ML predicted values and the DFT calculated values, while the SVM shows very weak correlation. The prediction accuracy of these CNNs for 5 ×5 supercell systems degrades a little (Fig. 3C). But the VCN network shows the prediction accuracy of > 90% for > 90% cases, which is the best among all three CNNs. The CCN has the lowest with prediction accuracy of > 90% for only ~ 50% cases. That is possible due to lack of training data for 5 × 5 supercell systems considering their much larger configuration space than 4 ×4 supercell systems. The ML predicted bandgaps still have strong linear correlation with the DFT calculated ones (Fig. 3D). They are still very impressive when compared with past ML materials (23, 41). We believe that as the size of training datasets increases, the accuracy would be further improved. Similar to the prediction results shown for 4 ×4 supercell systems, when predicting bandgaps of 5 ×5 supercell systems the SVM shows poor performance (Fig. 3B, D).

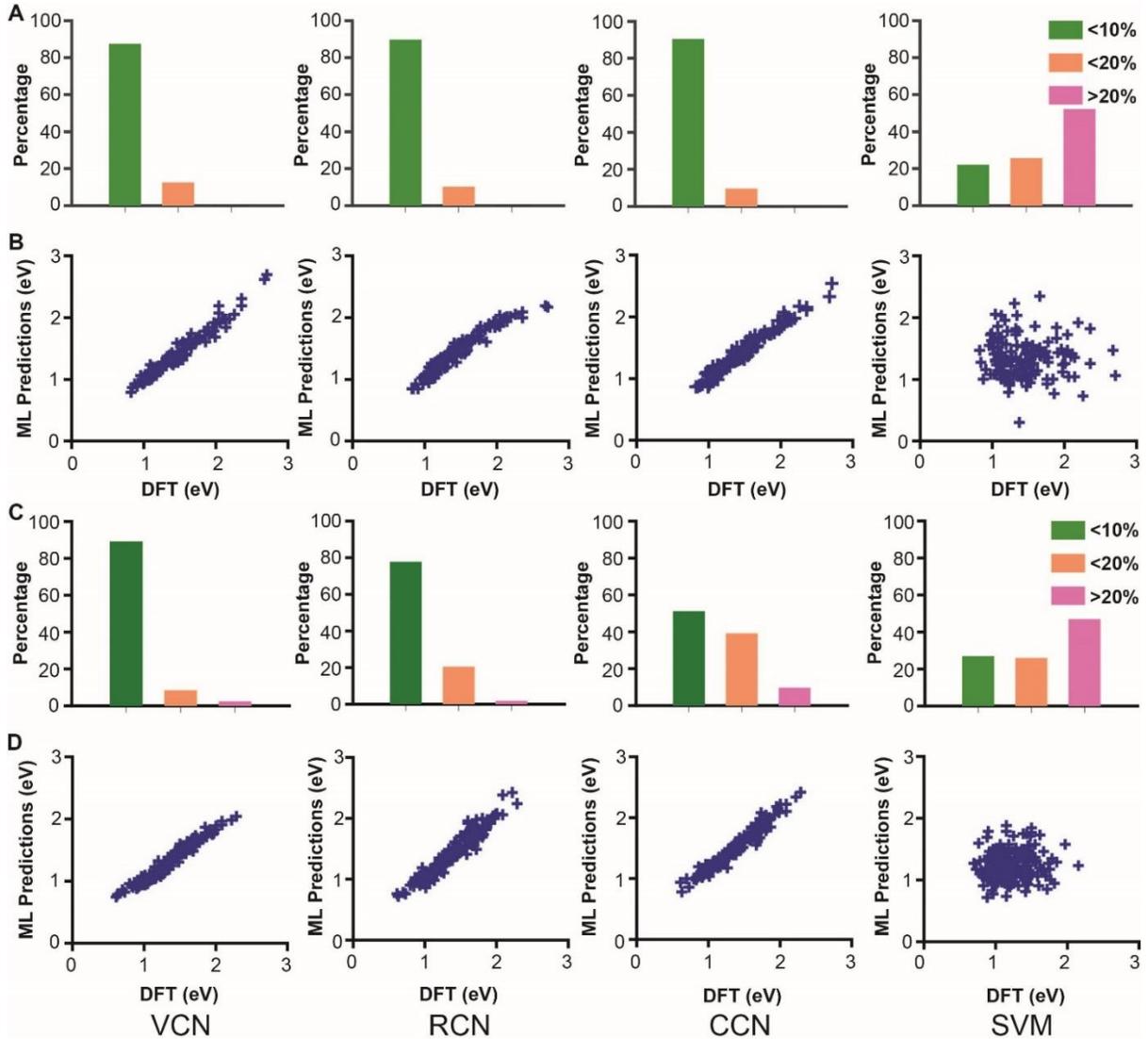

**Figure 3.** Prediction performance of four ML algorithms. (**A**, **C**) Error levels of ML predicted bandgaps for 4 ×4 supercell systems (**A**) and 5 ×5 supercell systems (**C**). The cases used for prediction are arranged by percentage of predicted cases showing prediction errors of $< 10\%$, $< 20\%$, and $> 20\%$. (**B**, **D**) ML predicted bandgaps versus DFT calculated values for 4 × 4 supercell systems (**B**) and 5 ×5 supercell systems (**D**).

In addition, other indicators of the prediction performance—the mean absolute error (MAE), root-mean-square error (RMSE) and explained variance ($R^2$)—are provided in Table 1. Meanwhile, the fractional error, $MAE_F$ and $RMSE_F$ were also calculated, with their definition is shown in Supplementary Notes 5. For the 4 ×4 supercell systems, all three CNNs show very low RMSE of ~ 0.1 eV for the predicted bandgaps. The corresponding fractional errors, $RMSE_F$, for all three CNNs are ~ 6%. For the 5 ×5 supercell systems, the RMSE values slightly increase to 0.1506 eV for the CCN, but it decreases to 0.0873 eV and 0.1044 eV when predicted by the VCN and RCN, further confirming the effectiveness of the VCN in predicting larger systems. The ultralow RMSE and $RMSE_F$ values show that these DL algorithms are more effective in predicting bandgaps of doped graphene than other material systems, such as double perovskites (41) which show a RMSE of 0.36 eV or inorganic crystals which show a RMSE of 0.51 eV.(23) This advantage is even more compelling if considering that: (i) our performance is rigorously evaluated with the newly generated systems that don't have any translational or symmetry equivalence with the training ones; (ii) the error maintains a low level when the relative size of the training data is significantly reduced for the 5 × 5 supercell systems. In contrast, the prediction accuracy from the SVM algorithm is much lower, showing much higher RMSEs of 0.33 and 0.51 eV for the 4 ×4 and 5 ×5 supercell systems, respectively. These values are closed to ones that are obtained from other non-CNN shallow learning algorithms such as the Kernel ridge regression(41) and the gradient boosting decision tree technique.(23) The $R^2$ is an indicator of correlation between the prediction and real values, which is considered as one of the most important metrics for evaluating the accuracy of the prediction models. Table 1 illustrates that the predicted bandgaps by all three CNNs have ~95% and > 90% relevance to the values calculated by the DFT for the 4 ×4 and 5 ×5 supercell systems, respectively. Among them, the RCN shows the best prediction results. In contrast, the SVM has a near zero $R^2$, indicating almost no relevance between the two. In summary, these results show that the CNNs are superior to non-CNN ML algorithms in predicting bandgaps of the topologically doped graphene.

**Table 1.** Statistics of predicted bandgaps by ML algorithms for 4 ×4 and 5 ×5 supercell systems.

|  | MSE (eV) | $MSE_F$ | RMSE (eV) | $RMSE_F$ | $R^2$ |
|---|---|---|---|---|---|
| 4 ×4 systems | | | | | |
| VCN | 0.0807 | 5.34% | 0.1030 | 6.47% | 0.9483 |
| RCN | 0.0776 | 5.12% | 0.1110 | 6.44% | 0.9426 |
| CCN | 0.0703 | 4.59% | 0.0945 | 5.83% | 0.9547 |
| SVM | 0.3990 | 27.6% | 0.3331 | 34.7% | 0.0029 |
| 5 ×5 systems | | | | | |
| VCN | 0.0693 | 5.33% | 0.0873 | 6.86% | 0.9622 |
| RCN | 0.0822 | 6.37% | 0.1044 | 8.06% | 0.9235 |
| CCN | 0.1351 | 11.3% | 0.1506 | 13.6% | 0.9469 |
| SVM | 0.2667 | 22.8% | 0.5120 | 29.5% | 0.0032 |

**2.4 Transfer learning: training and prediction**

As suggested from the prediction results shown in the 4 ×4 and 5 ×5 supercell systems, the prediction accuracy is decreased as the relative size of training data shrinks. Obtaining sufficient training data, such as bandgaps calculated by the DFT, can lead to unusually high cost especially as the system scales up. Such a problem imposes a major challenge in the application of machine learning to the materials science. To overcome this challenge, an emerging transfer learning (TL) has been proposed.(47) To conceptually validate the effectiveness of the TL in predicting bandgaps of larger systems we leveraged relatively larger datasets generated from the 4 ×4 and 5 ×5 supercell systems to improve models trained on more limited datasets generated in the 6 ×6 supercell systems. To do that, we built TL frameworks based on the CCN, RCN, and VCN. These networks were trained with the datasets previously used for training the 4 ×4 and 5 ×5 supercell systems together with 7200 new data points generated from the 6 ×6 supercell systems. The TL procedures for all CNNs are similar to the ones shown in Ref.,(48, 49) where all CNNs layers except the last fully-connected layer are set at a learning rate 10% of the original learning rate. The last layer is re-normalized and trained with the new dataset. Its learning rate is set to 1% of the original CNN networks.

Fig. 4A shows prediction errors of bandgaps for the 6 ×6 supercell systems by all three DL algorithms without the TL. Compared with the 4 ×4 and 5 ×5 systems they are significantly increased in all three categories of the prediction errors. The VCN shows > 90% prediction accuracy for only ~ 50% cases. The RCN performs the best among all three ones with near 55% cases reaching > 90% accuracy. The CCN shows the lowest percentage (i.e. ~ 45%) of cases within 10% prediction error and has 15% of the cases with > 20% error. Fig. 4B shows that the correlation between the predicted bandgaps and the DFT calculated ones is weaker for all three CNNs compared with the smaller systems, due to the much smaller percentage of sample size (only $1\times10^{-7}$ of all possibilities). The prediction accuracy can be notably boosted after they are combined the TL (Fig. 4C). The CCN with the TL performs the best, with > 60% cases achieving > 90% accuracy. The percentage of cases with > 20% error reduces to less than 10% from 20%. It is a significant improvement considering that increasing prediction accuracy becomes more and more difficult after a certain point. With the TL, the RCN and VCN also shows improved prediction accuracy. The percentage of the cases that show > 20% error decreases from 20% to 8% when using the RCN, while VCN decrease percentage of cases from 35% to 12%. The correlation between the predicted bandgaps and the DFT calculated ones becomes much stronger after using the TL methods (Fig. 4D). The statistics of the predicted bandgaps by the different DL algorithms with and without the TL for the $6 \times 6$ supercell systems is shown in Table 2. Overall, the TL boosts the prediction accuracy of all three DL algorithms in terms of reducing the MSE and RMSE. For instance, it helps to reduce the MSE of CCN from 0.1304 eV to 0.0910 eV and RMSE from 0.1634 eV to 0.1139 eV. Their $RMSE_F$ values are also reduced. This accuracy is comparable to the prediction accuracy show in the $5 \times 5$ supercell systems with the same DL algorithms. They are the lowest among the values predicted by all three CNNs used in this work. This demonstration of applying the TL to predict bandgaps of the topological doped graphene of larger size would pave new route to mitigating barriers for the machine learning in solving challenges of data scarcity faced in the material fields.

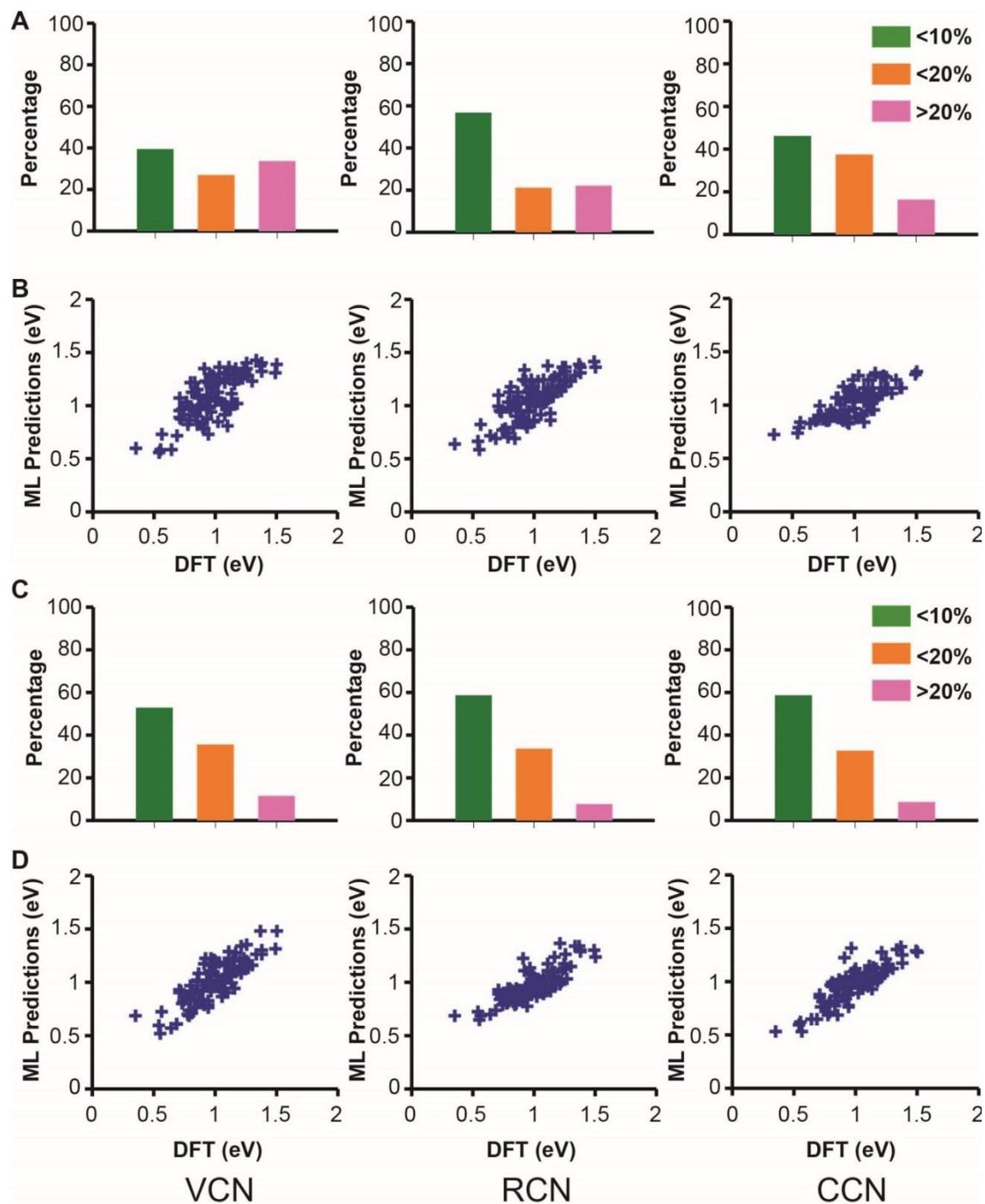

**Figure 4.** Prediction performance of three DL algorithms **(A, B)** before and **(C, D)** after transfer learning for 6 ×6 supercell systems. **(A, C)** Error levels of ML predicted bandgaps before TL (**A**) and after TL (**C**). The cases used for prediction are arranged by percentage of predicted cases

showing prediction errors of <10%, <20%, and >20%. (**B**, **D**) ML predicted bandgaps versus DFT calculated values before TL (**B**) and after TL (**D**).

**Table 2.** Statistics of predicted bandgaps by different DL algorithms without and with transfer learning for 6 ×6 supercell systems.

|     | MSE (eV) | $MSE_F$ | RMSE (eV) | $RMSE_F$ | $R^2$ |
| --- | --- | --- | --- | --- | --- |
| Without transfer learning ||||||
| VCN | 0.1493 | 16.4% | 0.1798 | 20.9% | 0.5135 |
| RCN | 0.1182 | 13.6% | 0.1529 | 19.4% | 0.5243 |
| CCN | 0.1304 | 14.5% | 0.1634 | 19.8% | 0.5771 |
| With transfer learning ||||||
| VCN | 0.1031 | 11.4% | 0.1237 | 15.9% | 0.6266 |
| RCN | 0.0960 | 10.5% | 0.1206 | 15.3% | 0.6463 |
| CCN | 0.0910 | 9.60% | 0.1139 | 12.6% | 0.6891 |

## Summary

In this work, we develop the DL models to predict the bandgaps of doped graphene with random dopant topologies. Three CNNs yield high-accuracy prediction results. The CNNs show superior performance to the non-CNN ML algorithms. The TL, leveraging the pre-trained network on small systems, boosts the prediction accuracy of three CNNs when predicting the bandgaps of large systems. The resulting MSE and RMSE of the predicted bandgaps of 6 × 6 systems by the CCN can be close to those of the 5 × 5 systems, but with a much smaller sampling ratio. Through this scientifically significant example, we have successfully illustrated the potential of artificial intelligence in studying 2D materials with topological dopant states, which would pave a new route to rational design of materials.

## Methods

**DFT calculation.** The *ab initio* DFT calculations were performed by QUANTUM ESPRESSO package (50). We have successfully employed DFT calculations to investigate the nitrogen doping in graphene and to predict a class of novel two-dimensional carbon nitrides.(51, 52) The ultra-soft projector-augmented wave (PAW) pseudopotential(53) was used to describe the interaction between the valence electrons and the ions. The Perdew-Burke-Ernzenhof (PBE) function was applied for the exchange-correlation energy.(54) The cutoff plane wave energy was set to 400 eV. The Monkhorst-Pack scheme(55) was applied to sample the Brillouin zone with a mesh grid from $12 \times 12 \times 1$ in the k-point for all the doped graphene. The graphene sheets were modeled as 2D matrixes. The matrix with all zero represents the intrinsic graphene, and the matrix with all one represents the intrinsic h-BN. The B-N pair doped graphene sheets were modeled by matrixes with 0 and 1 elements. The sizes of computed 2D sheets were $4 \times 4$, $5 \times 5$, and $6 \times 6$ supercells. The training datasets were the bandgaps of randomly generated 2D sheets with random concentration of B-N pair dopants. Specifically, the topologies of 2D graphene were generated by random uniform sampling. For example, for the $4 \times 4$ systems, we generated a set of random numbers between $0 \sim 2^{16}$. Then these decimal numbers were converted into binary numbers with 16 digits, which were further converted to 4 by 4 matrixes to represent samples with various dopant topologies.

## Acknowledgments

J. L. acknowledges financial support from University of Missouri-Columbia start-up fund, NASA Missouri Space Consortium (Project: 00049784), Unite States Department of Agriculture (Award number: 2018-67017-27880), US Department of Energy (Award number: DE-FE0031645) and National Science Foundation (Award numbers: DBI1759934 and IIS1763246). The computations were performed on the HPC resources at the University of Missouri Bioinformatics Consortium (UMBC), supported in part by NSF (award number: 1429294).

**Author contributions:** J. L. conceived the project. Y. D. designed descriptive system and performed the DFT calculations. C. H. W. built the CNNs and performed the training and prediction. J. L. and J. L. C. supervised the project. All authors contributed to the discussions and wrote the manuscript. **Competing interests:** There are no competing interests. **Materials & Correspondence:** Correspondence and material requests should be addressed to J. L. (LinJian@missouri.edu) and J. L. C. (chengji@missouri.edu). **Data availability:** All data needed to evaluate the conclusions in the paper are present in the paper and/or the Supplementary Materials. Additional data related to this paper may be requested from the authors.

# Supporting Information for

# Deep Learning Bandgaps of Topologically Doped Graphene

Yuan Dong[1,ξ], Chuhan Wu[2,ξ], Chi Zhang[1], Yingda Liu[2], Jianlin Cheng[2*], Jian Lin[1*]

[1]Department of Mechanical & Aerospace Engineering

[2]Department of Electrical Engineering & Computer Science, University of Missouri, Columbia, Missouri 65211, USA

[*]E-mail: LinJian@missouri.edu (J. L.) or chengji@missouri.edu (J. C.)
[ξ]Authors contributed equally to this work.

**Supplementary Text**

**Supplementary Note 1. Support Vector Machine (SVM)**

The SVM used in this work was developed by Corinna Cortes and Vapnik in 1993 (1), and evolved into support vector regression machines later (2). It was one of the hottest machine learning methods before the deep learning made a huge breakthrough recently. In this work, we first flatten the original input training data to a column vector (for example, the $4 \times 4$ data was flatten to be a $16 \times 1$ vector data). These column vectors were used as the SVM input. We selected Polynomial (3) as the regression kernel which can transform the data from low-dimension space to high-dimension space.

**Supplementary Note 2. VGG16 Convolutional Network (VCN)**

Convolutional neural networks (CNNs) (4), state-of-art deep learning methodologies which combine convolution layers and fully connected (FC) layers into one model, are used widely in the field of computer vision (5), natural language processing (6), self-driving cars (7), bioinformatics (8) and so on. For the raw 2D data, convolution can extract the feature not only from the element in the input data but also their neighbor elements' information. This is a feature detector that's useful in full perspectives of the 2D data. Convolution process converts them into feature map and transforms it to later layers. For instance, when applied to image data study, it means detecting the edge or textures of the images. In this study, the structure information of the doped graphene we use can be set as 2D data (Channel [might: 1]*Length*Width). Each pair of atoms inside its structure might infect its neighbor atoms so they collectively determine bandgaps of the whole structure. We firstly built a neural network by following the VGG16 structure, which was widely used in computer vision. This model has 16 convolution layers, one global-flatten layer, three FC layers and an output layer. We call it VGG16 convolutional network (VCN). This network was further modified into other two novel networks, residual convolutional network (RCN), and concatenate convolutional network (CCN). We will elaborate them in the following section.

## 2.1 Convolution layer

The convolution layer of CNN that it does convolution operation toward its input and output forms a 2D feature map. The convolution process equation is shown in the following.

$$(h_k)_{ij} = (W_k * x)_{ij} + b_k \quad (1)$$

$k$ is the index of the $k^{th}$ feature map. $W_k$ and $b_k$ are the weight and bias of $k^{th}$ feature map. $(h_k)_{ij}$ is the value of the output for the neuron in the $k^{th}$ feature map in position of (i; j).

## 2.2 Global Max-Pooling layer

This layer extracts the biggest element, which is also the most "important" and "significant" feature, from each channel of the previous layer. It offers the models the most convincing evidence to predict the bandgaps of the investigated systems. Since the feature map in each channel only outputs their biggest element, the output volume will "lost" one dimension from the input volume. Since this layer can convert 2D feature data into 1D vector feature data, it also acts as a "bridge layer" to link the previous convolution layer to the next Fully Connected Layer of the model.

### 2.3 Fully connected layer

The fully connected (FC) layer is a classical neural network layer where all the neurons from the previous layers are connected. It can be expressed as:

$$z_i = \sum_{k=1}^{N_1} w_{ki} x_k + b_i \qquad (2)$$

where $Z_i$ is the $i^{th}$ neuron's output, $N_1$ is the neurons contained in its previous layer, $w_{ki}$ is the weight of $k^{th}$ neuron from previous layer, and $b$ is the bias of the current layer.

### 2.4 Batch normalization layer

After each convolution and FC layer, there is a batch normalization layer. Batch normalization (9) is an useful process that it can erase the uncertainty of the hidden layer, and reduce the influence of internal covariate shift which changes the distribution of network activations during updating the parameters when training the network. Meanwhile it accelerates the training speed for the networks. The math of batch normalization is shown in the following:

$$\begin{align} \mu &= \frac{1}{m} \sum_i z^{(i)}, \\ \sigma^2 &= \frac{1}{m} \sum_i (z^{(i)} - \mu)^2, \\ Z_{norm}^{(i)} &= \frac{(z^{(i)} - \mu)}{\sqrt{\sigma^2 + \varepsilon}}, \\ \tilde{Z}^{(i)} &= \gamma \cdot Z_{norm}^{(i)} + \beta \end{align} \qquad (3)$$

Where $z^{(1)}, z^{(2)}, ..., z^{(m)}$ are the outputs of each layer; ε is introduced to avoid dividing by zero; β, γ are the learnable parameters of the model. After we replace $z^{(i)}$ with $\tilde{z}^{(i)}$ and push it into forward and backward propagation, the original hyper parameter bias $b^{(i)}$ will be replaced with $\beta$.

### 2.5 Activation function of exponential linear units (ELU)

We used exponential linear units (ELU) (10) as the activation function in each layer. The ELU is shown as follows when α is larger than zero.

$$f(x) = \begin{cases} x & \text{if } x > 0 \\ \alpha(\exp(x)-1) & \text{if } x \leq 0 \end{cases}, \quad f'(x) = \begin{cases} 1 & \text{if } x > 0 \\ f(x)+\alpha & \text{if } x \leq 0 \end{cases} \quad (4)$$

ELU tolerates the negative values when they are close to zero, thus it decreases shifting effect of the bias. The rectified linear unit (ReLU) activation function is used in previous ML works (11). However, in our network, the ReLU function could cause serious problem because it will always give zero if neuron units after ReLU activation converge to zero (called "dying ReLU" problem, http://cs231n.github.io/neural-networks-1/#actfun). So we choose to use ELU as our activation function and this problem is removed.

### Supplementary Note 3. Residual convolutional network (RCN)

The deep training of the neural network helps it to think deeper and can extract more details and important features from the training data. However, as the network becomes deeper and deeper, it will reach to a maximum point of performance, and thereafter encounter with degradation problem. The prediction accuracy gets saturated and degrades rapidly [3]. The residual network is introduced which converts each convolution layers to residual blocks [3]. It could help to solve the degradation problem. **Fig. S2** shows the structure of a residual block. This block has one convolution block and two identity blocks. The equation is shown as:

$$a^{[l+2]} = g(Z^{[l+2]} + a^{[l]}) = g(\omega^{[l+2]}a^{[l+1]} + b^{[l+2]} + a^{[l]}) \quad (5)$$

$Z^{[l]}$ : the input of $l^{th}$ activation layer
$\omega^{[l]}$ : the parameters of $l^{th}$ convolutional layer
$b^{[l]}$ : the bias of $l^{th}$ convolutional layer
$a^{[l]}$ : the input of residual block

When degradation problem occurs, the term of $(\omega^{[l+2]}a^{[l+1]} + b^{[l+2]}) = 0$, and this formula will become $a^{[l+2]} = g(a^{[l]})$, it can help to prevent the network from getting hurt by degradation and keep the network's strong robustness and its stable performance. Because the size of input is equal to the output of the residual block, the original network adds the parameters from $a[l]$ and $Z[l+1]$ in element wise.

**Supplementary Note 4. Concatenate Convolutional Network (CCN)**

The CCN combined featured advantages from GoogleNet (12) and DenseNet (13). Unlike RCN which "adds" feature maps in element-wise, this network concatenates the layer from input and output them pass though [l+α] activation layer together (**Fig. S3**). The equation of concatenate operation is:

$$\begin{aligned}\alpha^{[l+2]} &= C(g(Z^{[l+2]}), g(\alpha^{[l]}))_{axis=filter\_num} \\ &= C(g(\omega^{[l+2]}\alpha^{[l+1]} + b^{[l+2]}), g(\alpha^{[l]}))_{axis=filter\_num}\end{aligned} \quad (6)$$

where C(A,B) is the Concatenate function (Concatenate axis= Channels/filter numbers axis), $\alpha^{[l]}$ is the output of $l^{th}$ convolutional layer. When degradation problem occurs, it means

$$(\omega^{[l+2]}\alpha^{[l+1]} + b^{[l+2]}) = 0 \quad (7)$$

and the equation becomes

$$\alpha^{[l+2]} = C(g(0), g(\alpha^{[l]}))_{\text{axis=filter\_num}}$$
$$= C(0, g(\alpha^{[l]}))_{\text{axis=filter\_num}} \quad (8)$$

The input's original information still maintains in the output, so it can prevent the network from degradation. Meanwhile this process increases the volume and the trainable parameters of the data. They give more evidences and information for model to analyze and help it thinking "deep" and "thoroughly". Like the identity and convolution blocks of residual network, we introduced the small unit into the CCN: **Concatenation Block** (**Fig. S3B**). This block has three convolution layers: two normal convolution layers and one bottleneck layer inside of them. Bottleneck layer is a convolution layer with less filter number. Its filter size is 1×1 and convolution stride is 1. It reduces the number of parameters needed to update the neural without degrading the performance.

**Supplementary Note 5. Evaluation metrics**

The performance of prediction accuracy can be evaluated through many metrics. In this work we used coefficient of correlation (R), explained variance ($R^2$), Mean Absolute Error (MAE), and Root Mean Squared Error (RMSE). Their formulas are listed below:

$$R = \frac{\sum_{i=1}^{N}(X_i - \bar{X})(Y_i - \bar{Y})}{\sqrt{\sum_{i=1}^{N}(X_i - \bar{X})^2}\sqrt{\sum_{i=1}^{N}(Y_i - \bar{Y})^2}} \quad (9)$$

$$\text{MAE} = \frac{1}{N}|X_i - Y_i| \quad (10)$$

$$\text{RMSE} = \sqrt{\frac{1}{N}\sum_{i=1}^{N}(X_i - Y_i)^2} \quad (11)$$

where X is the DFT value and Y is the ML predicted value. $\bar{X}$ is the mean value of X, and analogously for $\bar{Y}$. N is the number of values used from comparison. The MAE and RMSE could be further normalized by X to express them as the fractional error.

$$\mathrm{MAE}_F = \frac{1}{N}\left|\frac{X_i - Y_i}{X_i}\right| \quad (12)$$

$$\mathrm{RMSE}_F = \sqrt{\frac{1}{N}\sum_{i=1}^{N}\left(\frac{X_i - Y_i}{X_i}\right)^2} \quad (13)$$

**Supplementary Figures**

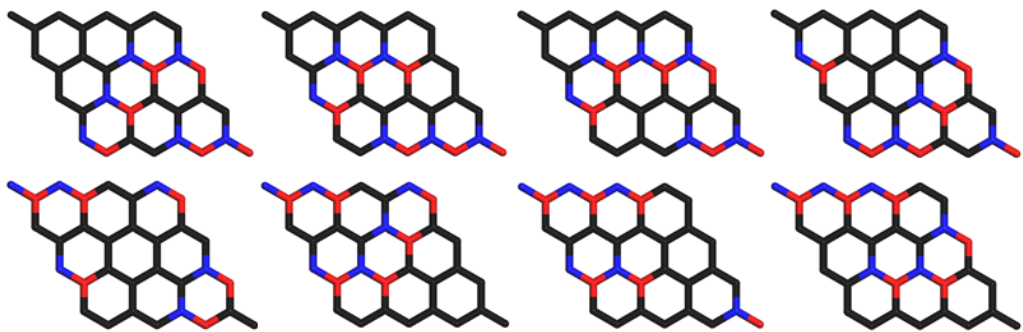

**Figure S1.** Examples of equivalent structures obtained by translating the particular structures along their lattice axis or inversion around their symmetry axis for 4 × 4 system.

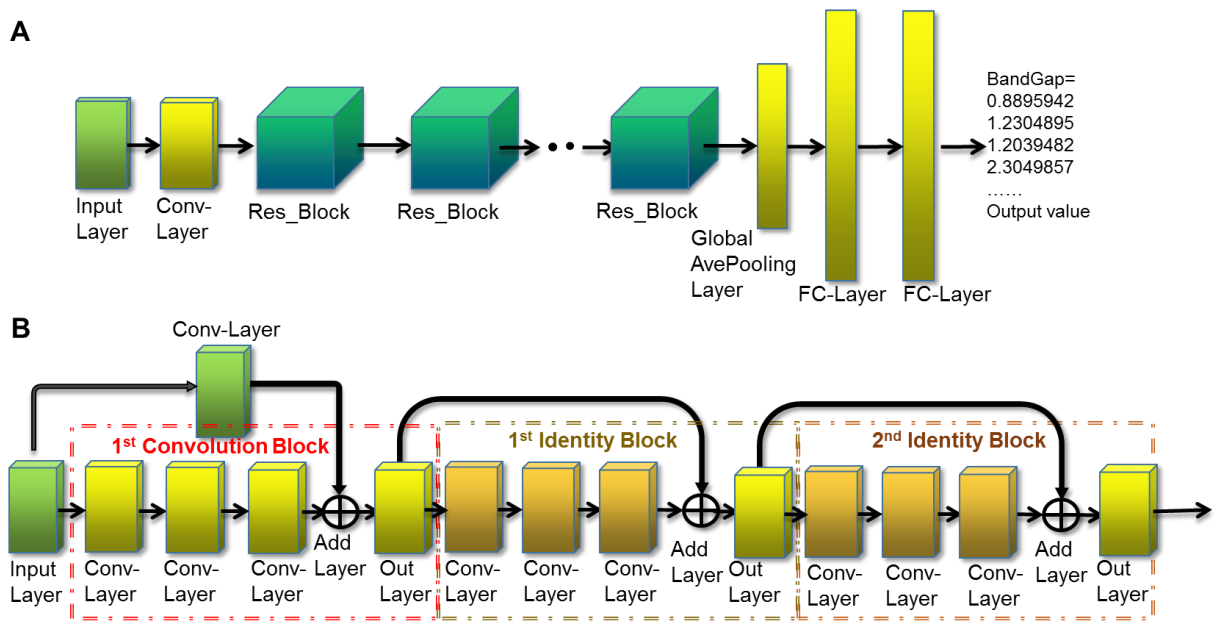

**Figure S2.** The structure of residual convolutional network (RCN). (A) The whole architecture of the RCN. (B) The detail about one example residual block (Res_Block).

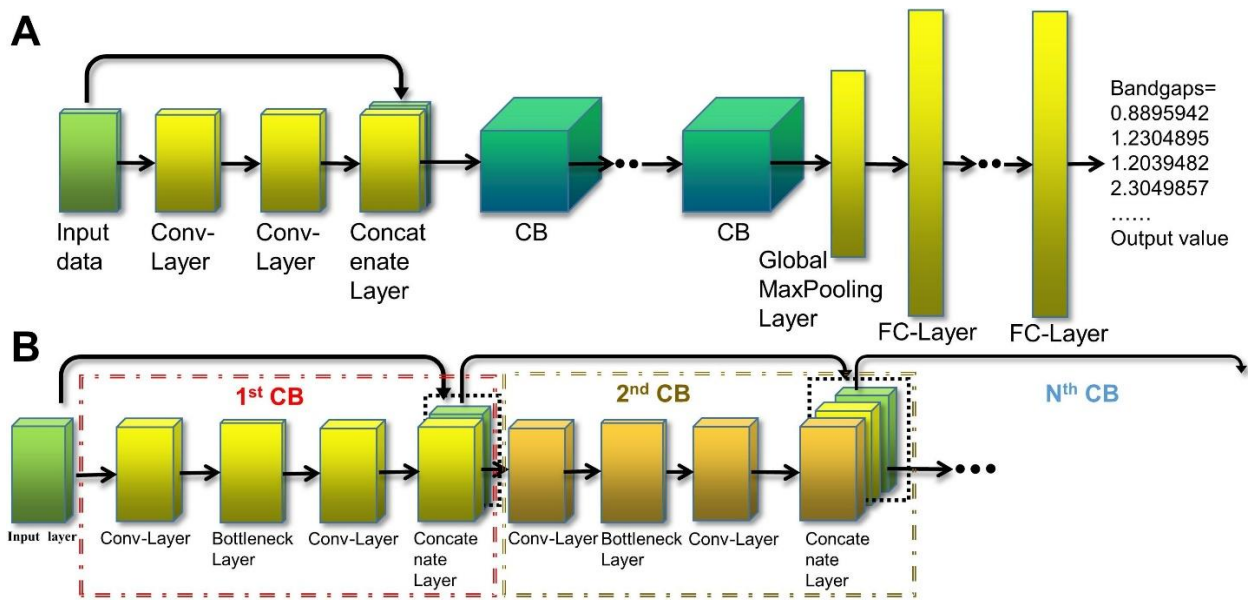

**Figure S3.** The structure of concatenate convolutional network (CCN). (A) Overall architecture of the CCN. (**B**) Details about the connection between two concatenation blocks (CBs) in the CCN.

**Supplementary Tables**

**Table S1.** Structure and hyperparameters of VGG16 convolutional network (VCN).

| |
|---|
| Input (Graphene Structure) |
| Convolution Layer 1 [filter number=50, filter size=(3,3), stride=(1,1), activation='elu'] Batch Normalization Layer 1 |
| Convolution Modules 2-12 Convolution Modules= {Convolution Layer [filter number=50, filter size=(3,3), stride=(1,1), activation='elu'] Batch Normalization Layer} |
| Global-Max Pooling Layer |
| Dropout Layer 1 (keep probability=0.6) |
| Fully-Collected Layer 1 (neuron unit=64, activation="elu") |
| Fully-Collected Layer 2 (neuron unit=64, activation="elu") |
| Fully-Collected Layer 3 (neuron unit=48, activation="elu") |
| Dropout Layer 2 (keep probability=0.7) |
| Output Layer (neuron unit=1, activation="elu") |
| Output (Bandgap Value) |

**Table S2.** Structure and hyperparameters of residual convolutional network (RCN).

| |
|---|
| Input (Graphene Structure) |
| Convolution Layer 1 [filter number=24, filter size=(3,3), stride=(1,1), activation='elu']<br>Batch Normalization Layer 1 |
| Residual Block 1:<br>{Convolution Block 1 [filter number 1=24, filter number 2=24, filter number 3=48]<br>Identity Block 1 [filter number 1=24, filter number 2=24, filter number 3=48]<br>Identity Block 2 [filter number 1=24, filter number 2=24, filter number 3=48]} |
| Residual Block 2:<br>{Convolution Block 2 [filter number 1=48, filter number 2=48, filter number 3=64]<br>Identity Block 3 [filter number 1=48, filter number 2=48, filter number 3=64]<br>Identity Block 4 [filter number 1=48, filter number 2=48, filter number 3=64]} |
| Residual Block 3:<br>{Convolution Block 3 [filter number 1=64, filter number 2=64, filter number 3=72]<br>Identity Block 5 [filter number 1=64, filter number 2=64, filter number 3=72]<br>Identity Block 6 [filter number 1=64, filter number 2=64, filter number 3=72]} |
| Residual Block 4:<br>{Convolution Block 4 [filter number 1=72, filter number 2=72, filter number 3=84]<br>Identity Block 7 [filter number 1=72, filter number 2=72, filter number 3=84]<br>Identity Block 8 [filter number 1=72, filter number 2=72, filter number 3=84]} |
| Global Average Pooling Layer |
| Fully-Collected Layer 1 (neuron unit=48, activation="elu") |
| Fully-Collected Layer 2 (neuron unit=24, activation="elu") |
| Output Layer (Neuron unit:1, activation="elu") |
| Output (Bandgap Value) |

**Table S3.** Structure and hyperparameters of convolution blocks inside residual convolutional network (RCN).

| Input | Input |
|---|---|
| Convolution Layer 1 (filter number 1, filter size=(1,1), stride=(1,1), activation='elu') | Convolution Layer 1 (filter number 3, filter size=(1,1), stride= (1,1), activation='elu') |
| Convolution Layer 2 (filter number 2, filter size=(3,3), stride=(1,1), activation='elu') | 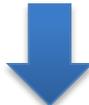 |
| Convolution Layer 3 (filter number 3, filter size=(1,1), stride=(1,1), activation='elu') | |
| Add Layer (activation='elu') ||
| Output Layer (Neuron unit=1, activation="elu") ||

**Table S4.** Structure and hyperparameters of identity blocks inside residual convolutional network (RCN).

| Input | Input |
|---|---|
| Convolution Layer 1 (filter number 1, filter size:(1,1), stride=(1,1), activation='elu') | 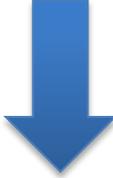 |
| Convolution Layer 2 (filter number 2, filter size:(3,3), stride=(1,1), activation='elu') | |
| Convolution Layer 3 (filter number 3, filter size:(1,1), stride=(1,1), activation='elu') | |
| Add Layer (activation='elu') ||
| Output Layer (Neuron unit=1, activation="elu") ||

**Table S5.** Structure and hyperparameters of concatenate convolutional network (CCN).

| Input (Graphene Structure) | Input (Graphene Structure) |
|---|---|
| Convolution Layer 1 [filter number=60, filter size=(3,3), stride=(1,1), activation='elu'] Batch Normalization Layer 1 | 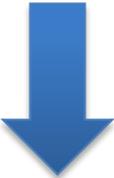 |
| Convolution Layer 2 [filter number=60, filter size=(3,3), stride=(1,1), activation='elu'] Batch Normalization Layer 2 | |
| Concatenate Layer (activation='elu') ||
| Concatenation Blocks 1-15 ||
| Global Max-Pooling Layer ||
| Drop Out Layer1 (keep probability=0.6) ||
| Fully-Collected Layer 1 (neuron unit=64, activation="elu") ||
| Fully-Collected Layer 2 (neuron unit=48, activation="elu") ||
| Fully-Collected Layer 3 (neuron unit=24, activation="elu") ||
| Drop Out Layer2 (keep probability=0.8) ||
| Output Layer (neuron unit=1, activation="elu") ||
| Output (Bandgap Value) ||

**Table S6.** Structure and hyperparameters of concatenation blocks inside concatenate convolutional network (CCN).

| Input | Input |
|---|---|
| Convolution Layer 1 (filter number=60, filter size=(3,3), stride=(1,1), activation='elu') | 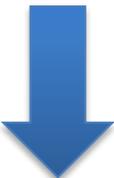 |
| Convolution Layer 2 ("Bottleneck Layer") (filter number=32, filter size=(1,1), stride=(1,1), activation='elu') | |
| Convolution Layer 3 (filter number=60, filter size=(3,3), stride=(1,1), activation='elu') | |
| Concatenate Layer (activation='elu') ||
| Output Layer (Neuron unit=1, activation="elu") ||

**Supplemental References**